\documentclass[aps,prl,twocolumn,superscriptaddress]{revtex4}
\usepackage{graphicx}
\usepackage{morefloats}
\usepackage{color}
\usepackage{amssymb}
\usepackage{float}
\usepackage{bm}
\usepackage{amsmath}
\usepackage{natbib}

\begin{document}

\newcommand{\ham}{\mathcal{H}} 
\newcommand{\Q}{\mathcal{Q}} 

\title{Spin caloritronic nano-oscillator}

\author{C.\,Safranski}
\thanks{These authors contributed equally to this work.}
\email{csafrans@uci.edu}
\author{I.\,Barsukov}
\thanks{These authors contributed equally to this work.}
\email{igorb@ucr.edu}
\affiliation{Department of Physics and Astronomy, University of California, Irvine, CA 92697, USA}
\author{H.\,K.\,Lee}
\affiliation{Department of Physics and Astronomy, University of California, Irvine, CA 92697, USA}

\author{T.\,Schneider}
\affiliation{Department of Physics and Astronomy, University of California, Irvine, CA 92697, USA}
\affiliation{Helmholtz-Zentrum Dresden--Rossendorf, Institute of Ion Beam Physics and Materials Research, Bautzner Landstrasse 400, 01328 Dresden, Germany}
\author{A.\,A.\,Jara}
\author{A.\,Smith}
\affiliation{Department of Physics and Astronomy, University of California, Irvine, CA 92697, USA}

\author{H.\,Chang}
\affiliation{Department of Physics, Colorado State University, Fort Collins, CO 80523, USA}

\author{K.\,Lenz}
\affiliation{Helmholtz-Zentrum Dresden--Rossendorf, Institute of Ion Beam Physics and Materials Research, Bautzner Landstrasse 400, 01328 Dresden, Germany}

\author{J.\,Lindner}
\affiliation{Helmholtz-Zentrum Dresden--Rossendorf, Institute of Ion Beam Physics and Materials Research, Bautzner Landstrasse 400, 01328 Dresden, Germany}

\author{Y.\,Tserkovnyak}
\affiliation{Department of Physics and Astronomy, University of California, Los Angeles, CA 90095, USA}

\author{M.\,Wu}
\affiliation{Department of Physics, Colorado State University, Fort Collins, CO 80523, USA}

\author{I.\,N.\,Krivorotov}
\email{ilya.krivorotov@uci.edu}
\affiliation{Department of Physics and Astronomy, University of California, Irvine, CA 92697, USA}

%\keywords{...}

\begin{abstract}
Energy loss due to ohmic heating is a major bottleneck limiting down-scaling and speed of nano-electronic devices, and harvesting ohmic heat for signal processing is a major challenge in modern electronics. Here we demonstrate that thermal gradients arising from ohmic heating can be utilized for excitation of coherent auto-oscillations of magnetization and for generation of tunable microwave signals. The heat-driven dynamics is observed in $\mathrm{Y_{3}Fe_{5}O_{12}/Pt}$ bilayer nanowires where ohmic heating of the Pt layer results in injection of pure spin current into the $\mathrm{Y_{3}Fe_{5}O_{12}}$ layer. This leads to excitation of auto-oscillations of the $\mathrm{Y_{3}Fe_{5}O_{12}}$ magnetization and generation of coherent microwave radiation. Our work paves the way towards spin caloritronic devices for microwave and magnonic applications.
\end{abstract}

\maketitle

Nano-devices based on control of magnetic damping by spin currents \cite{Flipse2014,choi2015}, such as spin torque memory and spin torque oscillators \cite{Kiselev2003,Mohseni,Rippard,Tamaru,Collet2016} (STOs), are at the forefront of spintronics research. In STOs, a  spin current injected into a ferromagnet acts as negative magnetic damping that cancels the positive damping of magnetization at a critical current density \cite{Kiselev2003,Mohseni,Rippard,Tamaru,Collet2016,Kajiwara,Demidov,hoffman2013,Liu2012,Duan2014}. Above the critical current, magnetization auto-oscillations are excited with an amplitude determined by nonlinearities of the magnetic system  \cite{Tiberkevich}.  By utilizing magnetoresistive effects, the auto-oscillations can be converted into microwave power.  STOs can also serve as sources of propagating spin waves for spin wave logic and nano-magnonic applications \cite{Akerman,Evelt}. STOs are typically metallic devices that rely either on spin-polarized electric currents \cite{Kiselev2003,Mohseni,Rippard} or on pure spin currents generated by the spin Hall effect \cite{Sinova,Demidov,Liu2012,Collet2016,Duan2014,Kajiwara}. Recently, an STO based on injection of pure spin Hall current into an insulating ferromagnet was demonstrated as well \cite{Collet2016}.

\begin{figure*}[pt]
\includegraphics[width=\textwidth]{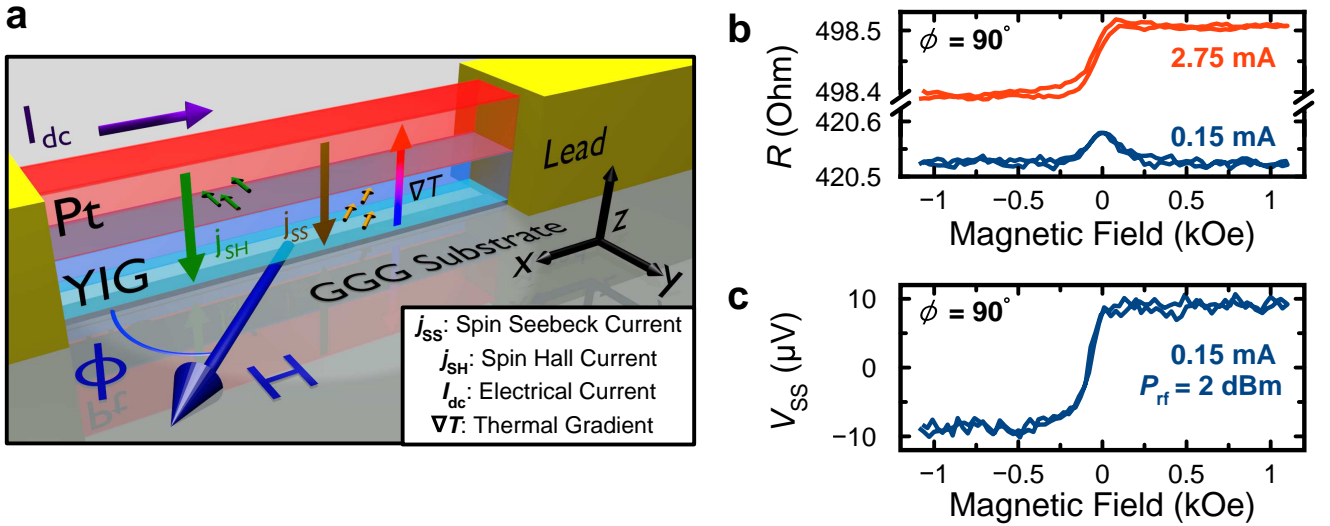}
\caption{\textbf{YIG/Pt nanowire magnetoresistance.} \textbf{a}~Sketch of the YIG/Pt nanowire spin torque oscillator. Arrows across the YIG/Pt interface illustrate the flow of spin Hall $j_{\mathrm{SH}}$ and spin Seebeck $j_{\mathrm{SS}}$ currents with corresponding smaller arrows representing the spin current polarization. The directions of the temperature gradient $\nabla T$ perpendicular to the YIG/Pt interface, the direct current $I_{\mathrm{dc}}$, and the magnetic field $H$ are depicted by arrows. \textbf{b}~Magnetoresistance $R$ of the YIG/Pt nanowire measured at low ($I_{\mathrm{dc}}=0.15$\,mA) and high ($I_{\mathrm{dc}}=2.75$\,mA) direct current bias for a magnetic field applied in the sample plane at the field angle $\phi=90^\circ$ with respect to the wire axis. \textbf{c}~ Spin Seebeck voltage $V_{\mathrm{SS}}$ induced in the nanowire by a large microwave current (microwave power $P_{\mathrm{rf}}=2$\,dBm) as a function of magnetic field.}
\label{setup}
\end{figure*} 

Spin angular momentum can be transferred from a nonmagnetic metal (N) to a ferromagnetic insulator (F) via inelastic spin flip scattering of a conduction electron at the F/N interface that generates a magnon in the F layer \cite{Evelt,Cunha,Wang2011,Bender2012,Hernandez2011}. The inverse process is responsible for the spin pumping effect whereby angular momentum of magnons in the F layer is converted into spin accumulation that drives spin current in the N layer \cite{Tserkovnyak2002b,Xiao}. At a non-zero temperature, these two processes give rise to a fluctuating spin current flowing across the F/N interface that time-averages to zero in thermal equilibrium \cite{Tserkovnyak2002b,Rezende2014}. When a thermal gradient is applied perpendicular to the F/N interface, a non-zero net spin current is established across the interface \cite{Ohnuma2015,Ohnuma2013,Rezende2014}. When the F layer is hotter than the N layer, a spin Seebeck current of magnons generated by the temperature gradient in the F layer flows towards the F/N interface and is converted at the interface into a pure spin current carried by conduction electrons in the N layer \cite{Jin2015,geprags2016,Wu2016,Xiao}. When the N layer is hotter than the F layer, the net spin current across the F/N interface reverses its direction resulting in angular momentum injection from the N layer into the F layer \cite{Cunha}. In this process, fluctuating spins of the conduction electrons of the hotter N layer generate a net flow of magnons into the colder F layer \cite{Rezende2011,Bender2016}, which increases the magnon density in the F layer above its equilibrium value \cite{Rezende2014,Ohnuma2015,Ohnuma2013,Bender2014,Bender2016}. 

The net spin current injected into the F layer by a temperature gradient across the F/N interface was theoretically predicted \cite{Slonczewski2010,Bender2012,Ohnuma2015,Ohnuma2013,Bauer,Rezende2014} and experimentally demonstrated \cite{Rezende2011,Cunha,MWU,Jungfleisch2013} to reduce the magnetic damping of the F layer magnetization. The action of such thermal spin current can be described in terms of an antidamping thermal spin torque \cite{Bender2016}, which can be called spin Seebeck torque. However, full cancellation of the F layer damping by the temperature gradient needed for the excitation of magnetic auto-oscillations has not been achieved.

In the following, we demonstrate operation of an STO driven by pure spin current arising from a temperature gradient across an F/N interface. This device realizes a major goal of spin caloritronics -- thermal energy harvesting for the manipulation of magnetization \cite{Bauer}.

\section*{Results}

\textbf{Sample description.} We study STOs based on pure spin current injection into a nanowire of an insulating ferrimagnet yttrium iron garnet $\mathrm{Y_{3}Fe_{5}O_{12}}$ (YIG). The 350\,nm wide, 15\,$\mu$m long nanowire devices are made by e-beam lithography and ion mill etching from $\mathrm{Gd_{3}Ga_{5}O_{12}}$(GGG substrate)/ YIG(23\,nm)/ Pt(8\,nm)  films grown by sputter deposition \cite{Chang}. Two  Al(4\,nm)/ Pt(2\,nm)/ Cu(15\,nm)/ Pt(2\,nm) leads with an inter-lead gap of 2.5\,$\mu$m are attached to the ends of the nanowire for application of an electric current along the wire length as shown in Fig.\,\ref{setup}a (and Supplementary Figure 1). Since YIG is an insulator, the electric current and the associated ohmic heat generation are confined to the Pt layer of the wire, which results in a large temperature gradient $\nabla T$ in the YIG film perpendicular to the YIG/Pt interface at high current densities (Supplementary Note 1 and Supplementary Figure 2).

\begin{figure*}[pt]
\includegraphics[width=\textwidth]{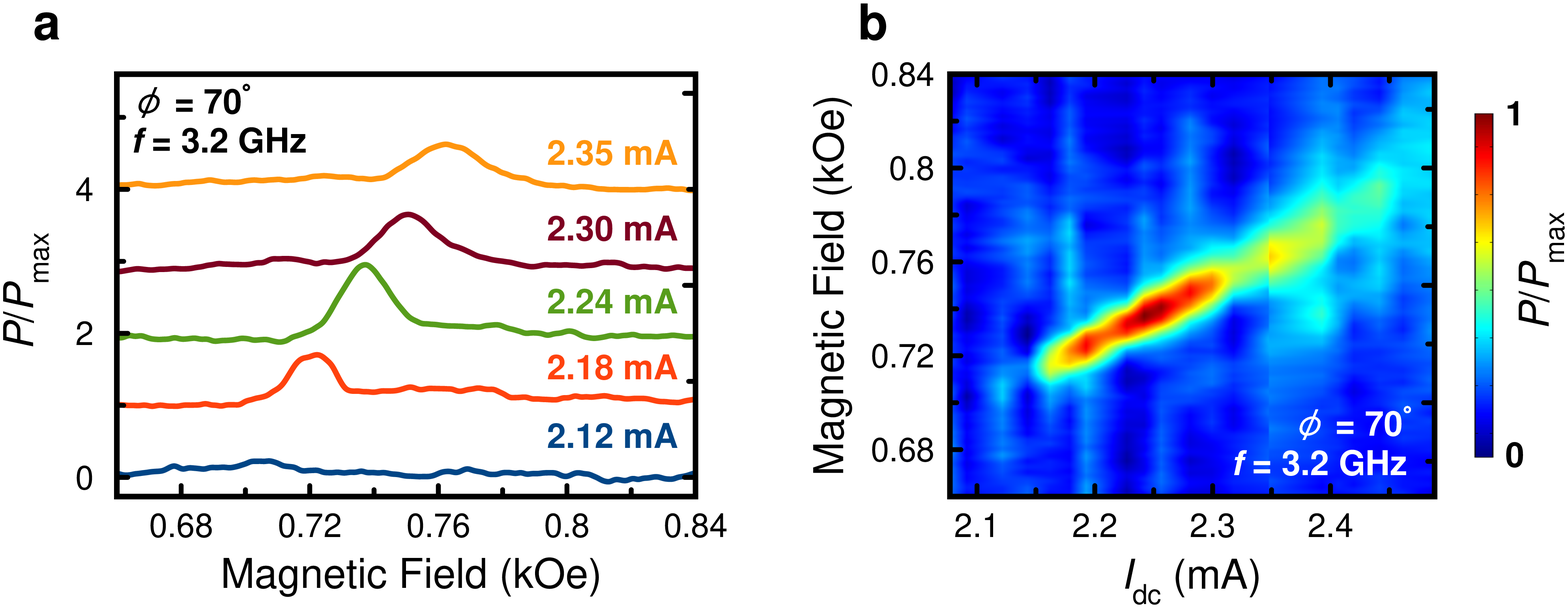}
\caption{\textbf{YIG/Pt nanowire microwave signal generation.} \textbf{a}~Spectra of normalized microwave power $P/P_{\mathrm{max}}$ generated by the nanowire at the frequency $f=3.2$\,GHz and magnetic field angle $\phi=70^\circ$ as a function of magnetic field at several direct current biases (vertically offset for clarity). \textbf{b}~Color plot of microwave power generated by the nanowire at 3.2\,GHz as a function of magnetic field and direct current bias.}
\label{setupb}
\end{figure*}

\textbf{Magnetoresistance measurements.} Figure\,\ref{setup}b shows the nanowire resistance as a function of magnetic field $H$ applied in the plane of the sample, perpendicular to the wire axis (in-plane field angle $\phi=90^\circ$). At a low direct bias current $I_{\mathrm{dc}}=0.15$\,mA, we observe negative magnetoresistance (MR) that saturates when $H$ exceeds the YIG nanowire's magnetic shape anisotropy field (0.2\,kOe). This data is well explained by spin Hall magnetoresistance (SMR) arising from spin Hall current in Pt flowing perpendicular to the YIG/Pt interface \cite{SMR,Miao}. At a  higher current bias $I_{\mathrm{dc}}=2.75$\,mA, the resistance of the wire saturates at different values for positive and negative $H$. This field-antisymmetric component of the MR arises from a large spin Seebeck current driven by $\nabla T$ and the inverse spin Hall effect in Pt \cite{Rezende2014,Uchida2010,Kikkawa2013, Vlietstra,Schreier2013,Guo2016, Schmid2013, Meier2013,Uchida2014}. As illustrated in Fig.\,\ref{setup}c, application of a large microwave current instead of $I_{\mathrm{dc}}$ also heats the Pt layer and results in a similar field-antisymmetric spin Seebeck voltage $V_{\mathrm{SS}}$ induced in the nanowire. This demonstrates that the field-antisymmetric MR in Fig.\,\ref{setup}b  arises from  $\nabla T$ rather than direct current  bias \cite{Schreier2013,Rezende2014}. All measurements presented in Figures 1--4 are made at the sample bath temperature of 140\,K chosen such that the nanowire temperature is near room temperature at the highest bias current (3\,mA) employed in our experiments (Methods, Supplementary Note 1 and Supplementary Figure 3).

\textbf{Microwave generation. } The MR data in Fig.\,\ref{setup}b,c suggest that pure spin currents driven by both the spin Hall effect and $\nabla T$ can flow across the YIG/Pt interface when a direct electric current is applied to the Pt layer. As discussed in the introduction,  these pure spin currents can apply antidamping spin torques to the magnetization of YIG \cite{Wang2011,Bender2016,Collet2016,Xiao}. If the negative effective damping due to these spin currents exceeds the positive magnetic damping of the YIG nanowire, GHz-range persistent auto-oscillatory dynamics of the YIG magnetization can be excited. Owing to the YIG/Pt nanowire MR, these auto-oscillations give rise to the sample resistance oscillations $\delta R_{\mathrm{ac}}$ and a microwave voltage  $ V_{\mathrm{ac}}\sim \delta R_{\mathrm{ac}}I_{\mathrm{dc}}$ generated by the sample at the frequency of the auto-oscillations. We detect the auto-oscillations of magnetization via measurements of the microwave power spectral density $P \sim V_{\mathrm{ac}}^2$ generated by the sample under direct current bias.

In contrast to conventional measurements of the microwave power emitted by STO as a function of frequency at fixed $H$ \cite{Duan2014}, we measure the emitted power at a fixed frequency (3.2\,GHz in Fig.\,\ref{setupb}a-b) as a function of the applied magnetic field $H$. As discussed in the Supplementary Note~2, this  STO characterization method greatly improves the signal-to-noise ratio in comparison to the conventional method and allows us to measure fW-scale microwave signals generated by nanomagnetic devices (Supplementary Figure 4). 

Our measurements of the microwave signal generated by the YIG/Pt nanowire reveal that full compensation of the  YIG layer damping by the spin Hall and the thermal spin currents can be achieved, and auto-oscillations of the YIG magnetization can be excited. Fig.\,\ref{setupb}a shows $P$ generated by the sample at 3.2\,GHz measured as a function of $H$ applied at $\phi=70^\circ$ for several values of $I_{\mathrm{dc}}$. These data reveal a sharp onset of the microwave emission from the sample for $I_{\mathrm{dc}}$ exceeding the critical value of 2.15\,mA. The peak of $P(H)$ for $I_{\mathrm{dc}}=2.15$\,mA is observed at $H=0.715$\,kOe, which is the resonance field of the lowest-frequency spin wave eigenmode of the YIG nanowire at this value of $I_{\mathrm{dc}}$, as shown in the next section. 

Fig.\,\ref{setupb}b demonstrates that the peak value of $P(H)$ shifts to higher magnetic fields with increasing  $I_{\mathrm{dc}}$. As discussed in the next section, this shift arises from reduction of the magnetization and magnetic shape anisotropy of the YIG wire caused by ohmic heating of the sample. The amplitude of the peak in $P(H)$ first increases with increasing $I_{\mathrm{dc}}$ reaching the maximum value $P_{\mathrm{max}} \approx 0.1$\,fW\,MHz$^{-1}$ at $I_{\mathrm{dc}}= 2.25$\,mA and then gradually decreases reaching the background noise level at $I_{\mathrm{dc}} \approx 2.5$\,mA for $\phi=70^\circ$. The integrated microwave power generated by the sample at $I_{\mathrm{dc}}=2.25$\,mA is estimated to be 6\,fW (Supplementary Note 3).

The field-frequency relation of the spin wave mode and thus the auto-oscillation frequency can be efficiently controlled via changing the nanowire width and thereby modifying its magnetic shape anisotropy \cite{Duan2015}. For the nanowire magnetized by a transverse magnetic field, the resonance field increases (the resonance frequency decreases) with decreasing nanowire width \cite{Duan2015}, as confirmed by our measurements of a 90\,nm wide YIG/Pt nanowire described in the Supplementary Note 2.

In order to gauge the relative contributions of the spin Hall and spin Seebeck currents to the excitation of the auto-oscillatory dynamics, we make measurements of the critical current for the onset of the auto-oscillations as a function of the in-plane magnetic field direction $\phi$. In-plane rotation of the YIG magnetization changes its direction with respect to the polarization of the spin Hall current, which leads to a $1/\sin \phi$ dependence of the critical current when  the spin Hall current is acting alone \cite{Collet2016}.  In contrast, the spin Seebeck current polarization is always collinear to the direction of the YIG magnetization. For $\nabla T$ shown in Fig.\,1a, the polarization is antiparallel to the magnetization of YIG, resulting in an antidamping spin torque \cite{Bender2016,Xiao,Ohnuma2013,Ohnuma2015}. The critical current due to the spin Seebeck current driving the magnetic precession is expected to be only weakly dependent on $\phi$  as long as the auto-oscillation frequency is fixed \cite{Bender2014} as in our measurements.

\begin{figure}[pt]
\includegraphics[width=0.5\textwidth]{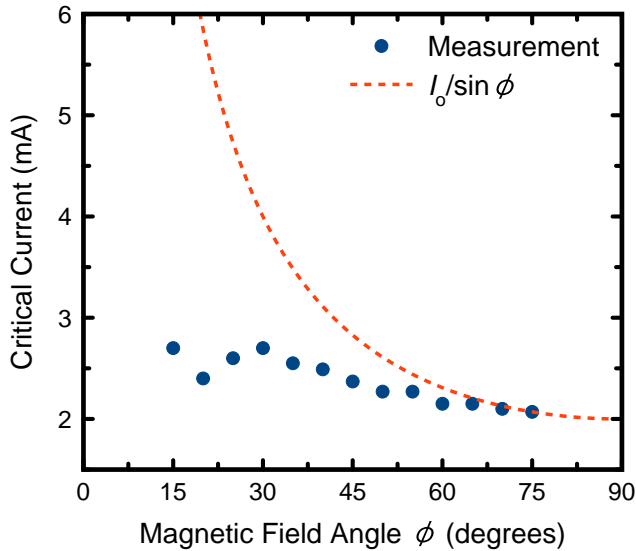}
\caption{\textbf{Angular dependence of the critical current.} Critical current for the onset of auto-oscillations as a function of in-plane magnetic field direction $\phi$. The line shows the expected behavior due to only  a spin Hall current (with a fitting parameter $I_{0}$) in the absence of a spin Seebeck current. }
\label{angular}
\end{figure} 

Figure\,\ref{angular} shows  the angular dependence of the critical current measured for our YIG/Pt nanowire in the range of $\phi$ from  $15^\circ$ to $75^\circ$.  For $\phi$ near  $0^\circ$ and $90^\circ$, the microwave power generated by the device is too small to be measured by our technique due to the weak angular dependence of MR near these angles. The measured angular dependence of the critical current is much weaker than the $1/\sin\phi$ dependence expected from the spin Hall current \cite{Collet2016,Duan2014} suggesting a significant contribution to antidamping from the spin Seebeck current. 

Current-driven auto-oscillations of magnetization were recently observed in microdisks of YIG/Pt bilayers at room temperature \cite{Collet2016}. In these structures, we estimate the temperature gradient across the YIG film at the critical current ($\approx0.033$\,K\,nm$^{-1}$) to be much weaker than in our devices ($\approx0.26$\,K\,nm$^{-1}$). This is because the nanowire geometry in our device strongly enhances heat confinement within the Pt layer (Supplementary Note 1). The measured angular dependence of the critical current in YIG/Pt microdisks closely followed the $1/\sin \phi$ dependence \cite{Collet2016} indicating that the auto-oscillations of the YIG magnetization in these samples with small $\nabla T$ were driven purely by spin Hall torque. Comparison of these results to the data in Fig.\,\ref{angular} lends further support to a large $\phi$-independent spin Seebeck antidamping in our YIG/Pt nanowire STOs. While quantitative fitting of the data in Fig.\,\ref{angular} is difficult because the temperature dependences of the spin Hall and spin Seebeck currents are not well understood, it is clear that auto-oscillations of magnetization in our YIG/Pt nanowires can be driven primarily by spin Seebeck current for a magnetic field direction near the nanowire axis. 

\begin{figure*}[pt]
\includegraphics[width=\textwidth]{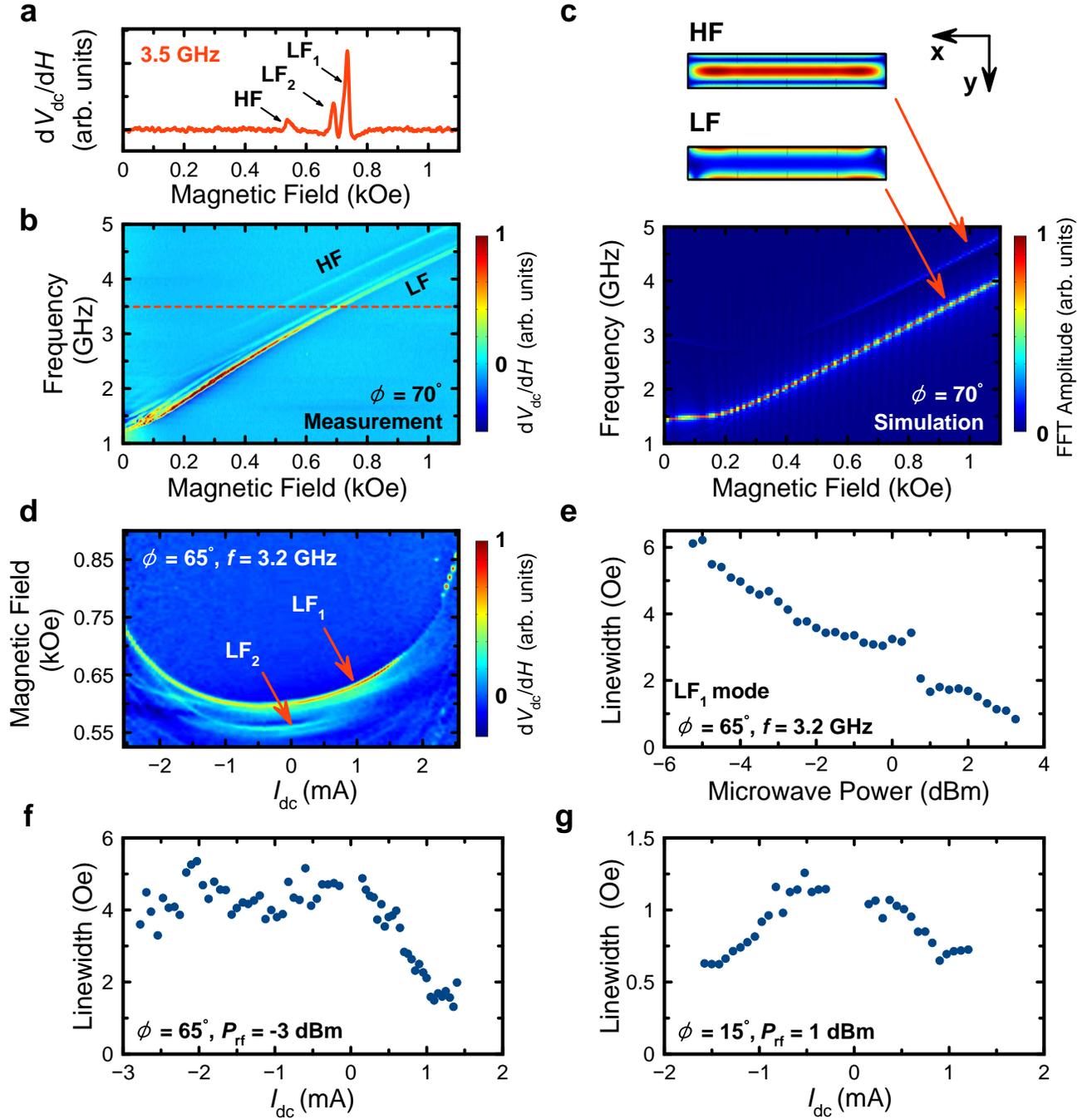}
\caption{\textbf{ ST-FMR measurements.} \textbf{ a}~A single spin-torque ferromagnetic resonance spectrum measured at a microwave frequency $f=3.5$\,GHz and magnetic field angle $\phi=70^\circ$. Low frequency (LF) and high frequency (HF) modes are observed. \textbf{ b} ST-FMR spectra of the YIG/Pt nanowire measured as a function of magnetic field and drive frequency at the field angle $\phi=70^\circ$.  \textbf{ c}~Micromagnetic simulation of the spin wave eigenmode spectra in the YIG/Pt nanowire with a top view of the spatial dependence of the LF and HF mode amplitudes. \textbf{ d}~ST-FMR spectra measured as a function of magnetic field and direct current bias current $I_{\mathrm{dc}}$ for  microwave power $P_{\mathrm{rf}}=-3$\,dBm. \textbf{ e}~Linewidth of the LF$_1$ measured as a function of the microwave drive power $P_{\mathrm{rf}}$. \textbf{ f}~Linewidth of the LF$_1$ mode as a function of direct current $I_{\mathrm{dc}}$ for $\phi=65^\circ$ and $P_{\mathrm{rf}}=-3$\,dBm. \textbf{ g}~Linewidth of the LF$_1$ mode as a function of direct current for $\phi=15^\circ$ and $P_{\mathrm{rf}}=1$\,dBm.}
\label{FMR}
\end{figure*} 

\textbf{Spin torque ferromagnetic resonance measurements.} To gain further understanding of the mechanisms leading to the excitation of the auto-oscillations, we measure the spectrum of spin wave eigenmodes of the nanowire by spin torque ferromagnetic resonance (ST-FMR) \cite{Sankey,tulapurkar2005}. In this method, a rectified voltage $V_{\mathrm{dc}}$ generated \cite{Chiba2014} by the sample in response to the applied microwave current is measured as a function of the drive frequency and external magnetic field. Resonances in $V_{\mathrm{dc}}$ are observed at the frequency and field values corresponding to spin wave eigenmodes of the system \cite{Ando2009}. To improve the sensitivity of the method, we modulate the applied magnetic field and measure $\frac{\mathrm{d}V_{\mathrm{dc}}}{\mathrm{d}H}$ \cite{Goncalves2013}.  In Figs.\,\ref{FMR}a,b, two spin wave eigenmodes of the YIG nanowire are clearly seen in the ST-FMR spectra measured at $I_{\mathrm{dc}}$=\,0. 

Patterning of the YIG film into the nanowire gives rise to lateral confinement of spin waves along the wire width, which results in quantization of the spin wave eigenmode spectrum \cite{Duan2014,Duan2015}. The resonances observed in ST-FMR spectra in Figs.\,\ref{FMR}a,b arise from the low frequency sector of this quantized spin wave mode spectrum. Precise identification of the modes measured by the ST-FMR spectroscopy can be accomplished via comparison of the ST-FMR data in Fig.\,\ref{FMR}b to the spin wave eigenmode spectra obtained from micromagnetic simulations \cite{MuMax3_2014}  (Supplementary Note 4). Fig.\,\ref{FMR}c shows the dependence of the simulated eigenmode frequencies on magnetic field for $\phi = 70^\circ$. Similar to the ST-FMR data in Fig.\,\ref{FMR}b, two eigenmodes are present in the simulations with their frequencies being close to those measured by ST-FMR. The spatial profiles of the amplitudes of these modes displayed in Fig.\,\ref{FMR}c for $\phi=70^\circ$ show that the high frequency (HF) mode is a standing spin wave with two nodes in the direction perpendicular to the wire axis. For the low frequency (LF) mode, the amplitude maxima lie near the wire edges \cite{Duan2014}. The LF mode exhibits fine splitting (labeled as LF$_1$ and LF$_2$ in Fig.\,\ref{FMR}a), which arises from geometric confinement \cite{Nembach2011,Hamadeh2014} of the mode along the wire length \cite{Duan2015,Duan2014}. 

Figure\,\ref{FMR}d shows ST-FMR spectra of the wire measured  at 3.2\,GHz and $\phi=65^\circ$ as a function of the direct bias current. At currents below the critical, the resonance fields of the LF$_1$ and LF$_2$ modes exhibit a current-induced shift. This shift has a quadratic component from the reduction of the YIG magnetization by ohmic heating, as established by previous studies of magnetic nanodevices \cite{Hamadeh2014,Petit2007}. The linear component of the frequency shift is due to the spin Hall current \cite{Demidov2011}.  At the critical current for the onset of auto-oscillations (2.2\,mA), the LF$_1$ mode resonance field is equal to the field at which auto-oscillations of magnetization are observed for this measurement frequency and direction of $H$. This demonstrates that the observed auto-oscillatory dynamics arises from the LF$_1$ mode.

Quantitative measurements of the spin wave mode intrinsic linewidth (Supplementary Note 5) by ST-FMR in our samples present significant difficulties as illustrated in Fig.\,\ref{FMR}e, which shows the linewidth of the LF$_1$ mode as a function of microwave drive power $P_{\mathrm{rf}}$. The observed linewidth decrease with increasing $P_{\mathrm{rf}}$ cannot be explained by a nonlinear lineshape distortion that is known to increase the linewidth with increasing power \cite{Sankey}. However, it is consistent with the antidamping action of the spin Seebeck current arising from microwave-induced heating of the Pt layer. Much lower values of the drive power needed for measuring the intrinsic linewidth (i.\,e. free from both spin Seebeck and nonlinear contributions) do not produce a measurable ST-FMR signal due to the small MR of the samples (Fig.\,\ref{setup}b). Therefore, ST-FMR measurements of the linewidth presented below are invasive. Nevertheless they give important qualitative information on the relative contributions of the spin Hall and spin Seebeck currents to the current-induced antidamping torque.

\section*{Discussion}

The linewidth of the  spin wave eigenmodes below the critical current in a spin Hall oscillator decreases with $I_{\mathrm{dc}}$ for one (antidamping) current polarity and increases for the other (damping) polarity because the damping-like component of the spin Hall torque is linear in $I_{\mathrm{dc}}$ \cite{Collet2016,Sinova,Petit2007,Hamadeh2014}. The dependence of the LF$_1$ mode linewidth on $I_{\mathrm{dc}}$ in our YIG/Pt devices exhibits a radically different behavior illustrated in Fig.\,\ref{FMR}f. The linewidth measured at $P_{\mathrm{rf}}=-3$\,dBm decreases for $I_{\mathrm{dc}}>0$ but remains nearly constant for $I_{\mathrm{dc}}<0$. These data can be explained via a significant antidamping spin Seebeck torque \cite{Rezende2014,Bender2016} driven by $\nabla T$. Indeed, both direct current polarities give rise to the same degree of ohmic heating in Pt and therefore to the same spin Seebeck current acting as antidamping \cite{MWU,Jungfleisch2013,Rezende2014,Rezende2011,Ohnuma2015,Ohnuma2013}. The rapid decrease of the linewidth with positive current at $\phi=65^\circ$ shown in Fig.\,\ref{FMR}f is due to the combined action of the antidamping torques from spin Hall and spin Seebeck currents. At a negative current bias, the spin Hall torque acts as positive damping while the spin Seebeck torque acts as antidamping. The competition between these two torques results in a weak variation of the linewidth with current for $I_{\mathrm{dc}}<0$ and $\phi=65^\circ$.

Application of the magnetic field nearly parallel to the nanowire axis ($\phi=15^\circ$) allows us to further separate the action of the spin Seebeck torque from that of the spin Hall torque. In this configuration, the polarization of the spin Hall current and magnetization are nearly perpendicular, resulting in a negligibly small damping/antidamping spin Hall torque \cite{Liu2012}. Thus, the $\phi$-independent antidamping torque from the spin Seebeck current \cite{MWU,Bender2016,Jungfleisch2013,Rezende2014} is expected to dominate the magnetization dynamics. ST-FMR measurement of the LF$_1$ mode linewidth as a function of current bias is shown in Fig.\,\ref{FMR}g. For this measurement we must employ a higher microwave drive ($P_{\mathrm{rf}}=1$\,dBm) due to the reduced ST-FMR signal in a nanowire magnetized close to its axis.  The data illustrate that we indeed observe the linewidth decrease with increasing current bias for both current polarities, as expected for the dominant antidamping from the spin Seebeck current and negligible damping/antidamping from the spin Hall current.  The linewidth decrease with increasing current bias in Fig.\,\ref{FMR}f and Fig.\,\ref{FMR}g cannot be explained by a temperature dependence of damping \cite{Haidar2015} since we find the linewidth to be nearly constant in the temperature range from 140\,K to 300\,K (Supplementary Note 6 and Supplementary Figure 5).

A microscopic mechanism of the antidamping action of the spin Seebeck current in F/N bilayers was theoretically discussed in Ref.~\cite{Bender2016}. In this mechanism,  a spin Seebeck current driven from the N into the F layer by $\nabla T$ generates a non-equilibrium population of incoherent magnons in the F layer. Nonlinear magnon scattering transfers angular momentum from this incoherent magnon cloud to a low-frequency coherent spin wave mode and thereby reduces the effective damping of the mode. In this picture, the auto-oscillatory spin wave dynamics above the critical current can be viewed as bosonic condensation \cite{Demokritov} of non-equilibrium incoherent magnons generated by  a thermal spin current into a coherent low-frequency spin wave mode \cite{Bender2012,bozhko2015}. While the order of magnitude of the threshold thermal bias measured in our experiment is reasonable, according to the theory (where the threshold temperature difference between magnons in YIG and electrons in Pt is set by the frequency of the auto-oscillating mode), the quantitative details depend on the interplay of the magnon-magnon, magnon-phonon, and magnon-electron scatterings. Future efforts are called upon to clarify the relative importance of these in YIG/Pt heterostructures.

In conclusion, we observe current-driven auto-oscillations of the magnetization in YIG/Pt bilayer nanowires. Measurements of the angular and current bias dependence of antidamping spin torques in this system reveal that the auto-oscillatory dynamics are  excited by a combination of spin Hall and spin Seebeck currents. We show that the spin Seebeck current resulting from ohmic heating of Pt can be the dominant drive of auto-oscillations of the YIG magnetization. Our measurements demonstrate that ohmic heating can be utilized for generation of tunable microwave signals and coherent spin waves. While the output power of the YIG/Pt STO studied here is low, we expect that the spin Seebeck drive mechanism demonstrated here can be utilized in other types of STOs with higher magnetoresistance and output microwave power. Our results pave the way towards spin caloritronic devices based on ohmic heat harvesting. 

%\textbf{\cite{Tserkovnyak2002b,Kajiwara}. \cite{Uchida2010,Bauer,Bender2012,Rezende2014,Ohnuma2015}.  \cite{Jin2015,Uchida2014,geprags2016,Guo2016,Schmid2013,Meier2013,Wu2016}  \cite{Slonczewski2010,Bender2016,Rezende2011}  \cite{Bender2012, MWU,Cunha, Jungfleisch2013}.}

\section*{Methods}

\textbf{Experimental technique.} All measurements reported in this study are performed in a continuous flow helium cryostat, where the sample is surrounded by a helium gas injected into the sample space via a needle valve at the bottom of the sample space. The temperature of the helium gas can be controlled in the range from 4.2\,K to 300\,K via a feedback loop using a heater and a thermometer near the needle valve. The sample is attached to a custom made brass sample holder placed near the helium gas injection port, and the contact pads of the YIG/Pt nanowire device are electrically connected to a short coplanar waveguide (CPW) via 4\,mm long aluminum wire bonds. The brass sample holder is electrically connected to the ground of the CPW, and the central conductor of the CPW is soldered to a microwave K-connector. The K-connector of the sample holder is connected to the microwave electronics outside of the cryostat (Supplementary Figure~4a) via a 1\,m long cryogenic microwave cable. The frequency-dependent microwave signal attenuation/amplification of the entire microwave circuit from the sample holder to the spectrum analyzer is characterized using a microwave network analyzer, and the reported  power levels are corrected for this loss/amplification. An electromagnet on a rotating stage placed outside the cryostat allows for application of a magnetic field up to 4\,kOe at an arbitrary direction within the sample plane.

The microwave signal generated by the YIG/Pt nanowire oscillator is low-level (power spectral density below 1\,fW\,MHz$^{-1}$) because magnetoresistance of the sample is small. In order to reliably measure the spectral properties of such low-level signals, we develop an ultra-sensitive technique for detection of microwave signals generated by spin torque oscillators (STOs). This technique improves the signal-to-noise ratio over the conventional technique used for measurements of STO microwave emission spectra by two orders of magnitude. The key feature of this technique is the harmonic modulation of the applied magnetic field and lock-in detection of the emitted microwave power at the modulation frequency, which greatly diminishes the influence of non-magnetic noise on the measured signal (Supplementary Note 2). 

\textbf{Data availability.} All data supporting the findings of this study are available within the article and the Supplementary Information file, or are available from the corresponding author on reasonable request.

\section*{References}

%\bibliographystyle{naturemag}
%\bibliography{main_final_cited}

\section*{Acknowledgments} 
This work was primarily supported as part of the Spins and Heat in Nanoscale Electronic Systems (SHINES), an Energy Frontier Research Center funded by the U.S. Department of Energy, Office of Basic Energy Sciences (BES) under Award No. DE-SC0012670. Y.T. acknowledges support by the U.S. Department of Energy, Office of Basic Energy Sciences under Award No. DE-SC0012190 (theory) and the Army Research Office under Grant No. W911NF-14-1-0016 (modeling). The work of A.A.J. (sample nanofabrication) was supported by the U.S. Department of Energy, Office of Basic Energy Sciences under Award No. DE-SC0014467.

\section*{Author contributions}
H.C. deposited the magnetic multilayers. C.S., A.A.J., and A.S. carried out the nanofabrication of STOs. C.S., H.K.L, and I.B. performed the characterization of the samples. T.S., K.L. and J.L. performed micromagnetic simulations. C.S. and I.B. developed the microwave measurement techniques used in this work. I.N.K., Y.T. and M.W formulated the problem and designed the experiment. I.N.K. managed the project. All authors analyzed the data and co-wrote the paper. C.S. and I.B. contributed equally to this work.

\section*{Competing financial interests} 
The authors declare no competing financial interests.

%\section*{Correspondence}
%Correspondence and requests for materials should be addressed to Ilya Krivorotov (email: ikrivoro@uci.edu).
\clearpage
\widetext

\begin{center}
\textbf{\large Supplemental Information: Spin caloritronic nano-oscillator}
\bigskip
\end{center}
\twocolumngrid
%%%%%%%%%% Merge with supplemental materials %%%%%%%%%%
%%%%%%%%%% Prefix a "S" to all equations, figures, tables and reset the counter %%%%%%%%%%
\setcounter{equation}{0}
\setcounter{figure}{0}
\setcounter{table}{0}
\setcounter{page}{1}
%\documentclass[aps,prl,twocolumn,superscriptaddress]{revtex4}
%\makeatletter
\renewcommand{\theequation}{S\arabic{equation}}
\renewcommand{\thefigure}{S\arabic{figure}}
\renewcommand{\bibnumfmt}[1]{[S#1]}
\renewcommand{\citenumfont}[1]{S#1}

\renewcommand{\figurename}{Supplementary Figure}
\renewcommand\refname{Supplementary References}
\def\bibsection{\section*{\refname}} 

\begin{figure*} 
\includegraphics[width=.6\textwidth]{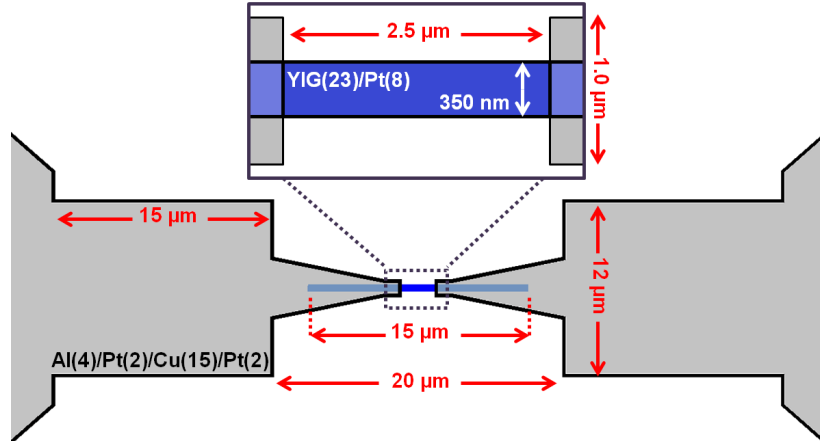}
\caption{\textbf{Nanowire device sketch.} Sketch of the YIG/Pt bilayer nanowire device with two Al/Pt/Cu/Pt leads. The nanowire depicted in blue color is partially covered by the electrical leads depicted in grey. The expanded view panel shows the active region of the nanowire. Lateral dimensions as well as nanowire and lead material stacks including layer thicknesses in nanometers are shown.}
\label{DevPat}
\end{figure*}

\begin{figure*} 
\includegraphics[width=.7\textwidth]{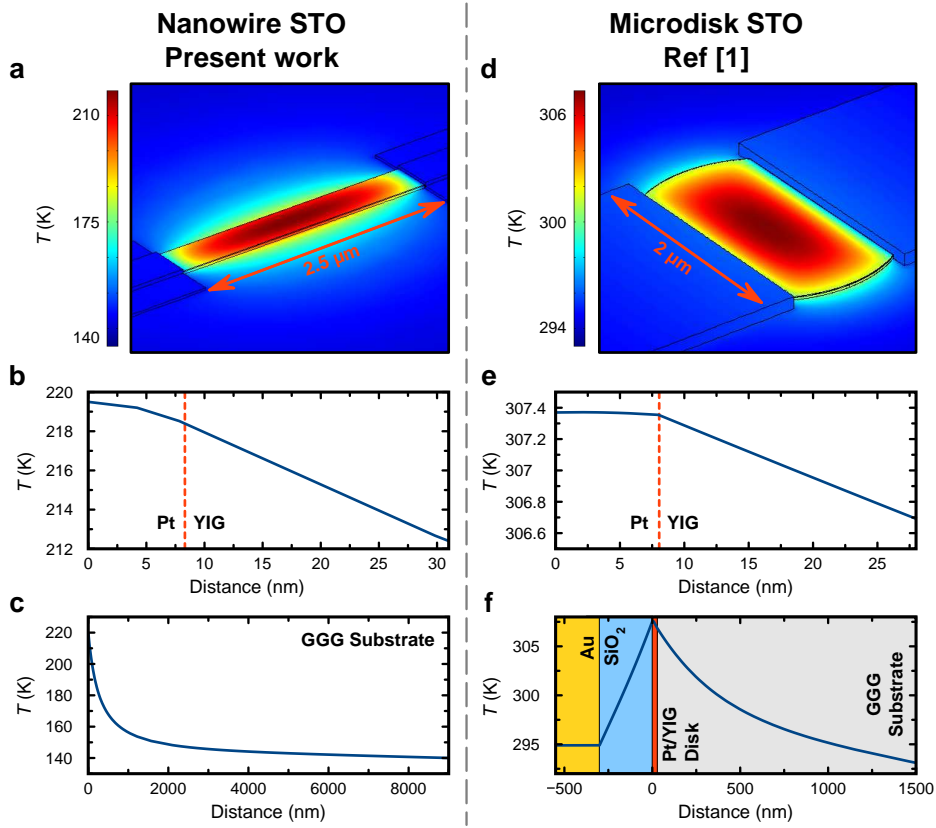}
\caption{\textbf{Ohmic heating and temperature gradients.} \textbf{ a}~COMSOL simulation of temperature distribution due to ohmic heating in a YIG(23\,nm)/Pt(8\,nm) nanowire spin torque oscillator (STO) at $I_{\mathrm{dc}}=$ 2.5\,mA and 140\,K bath temperature. \textbf{ b}~Temperature depth profile in the center of the nanowire.\textbf{ c}~Temperature depth profile within the GGG substrate under the nanowire center. \textbf{ d}~COMSOL simulation of temperature distribution due to ohmic heating in a YIG(20\,nm)/Pt(8\,nm) microdisk studied in Ref.\,\onlinecite{Collet2016} at $I_{\mathrm{dc}}=$ 7.4\,mA and 293\,K bath temperature.\textbf{ e}~Temperature depth profile in the center of the microdisk.\textbf{ f}~Temperature depth profile in the SiO$_2$/Au  overlayer above the microdisk and in the GGG substrate below the microdisk center.}
\label{1}
\end{figure*}

\begin{figure} 
\includegraphics[width=.42\textwidth]{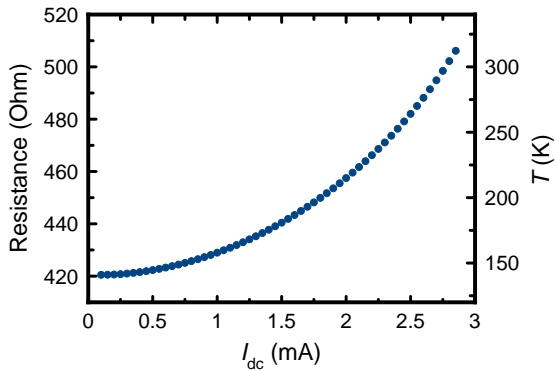}
\caption{\textbf{Pt nanowire resistance as a function of direct current bias.} The right ordinate axis shows the corresponding wire temperature that was obtained from measurements of the wire resistance as a function of bath temperature at small bias current.}
\label{2}
\end{figure}

\begin{figure*} 
\includegraphics[width=0.8\textwidth]{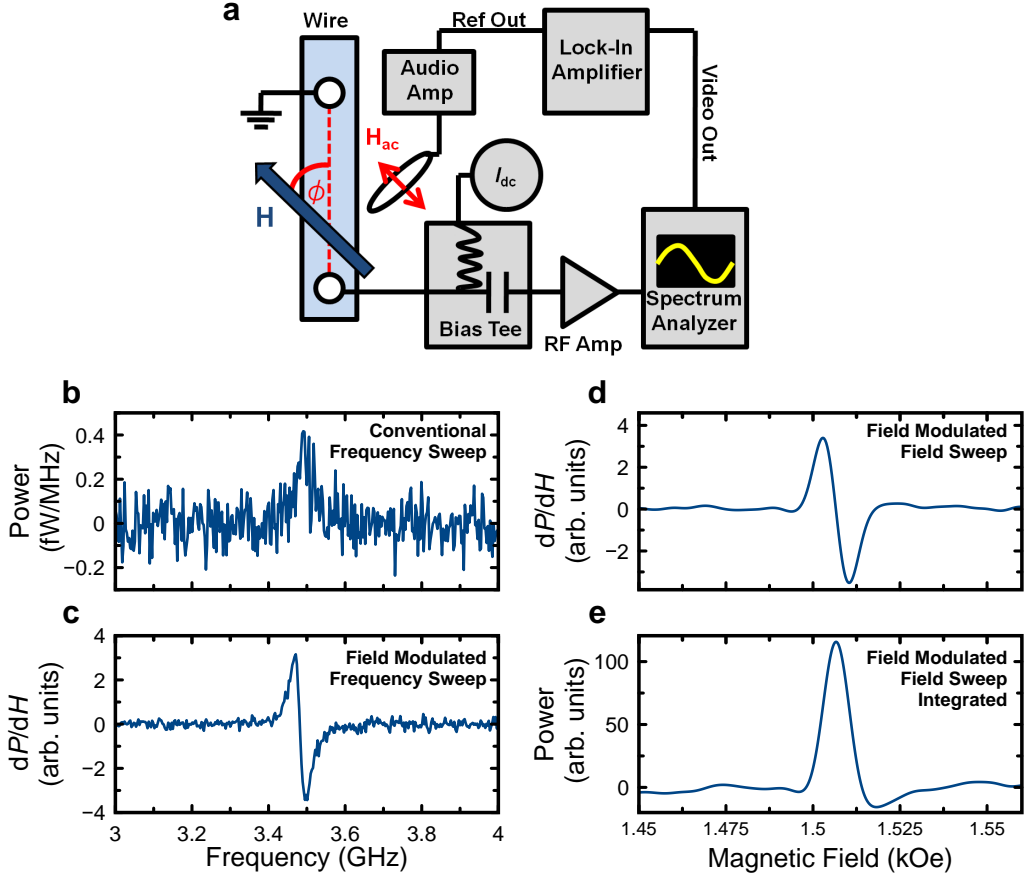}
\caption{\textbf{Microwave emission measurements.} \textbf{ a}~Schematic of the experimental setup for measurements of spin torque oscillator (STO) microwave emission with magnetic field modulation. The nanowire with magnetic field $H$ applied at an angle $\phi$ is supplied with a direct current $I_{\mathrm{dc}}$ via a bias tee. The microwave signal is amplified and detected using a spectrum analyzer. Its video-out signal is processed with a lock-in amplifier. The reference-out signal is used to generate a modulation field $H_{\mathrm{ac}}$ using an audio amplifier. \textbf{b-e} Microwave emission spectra from a 90\,nm wide YIG(23\,nm)/Pt(8\,nm) nanowire STO with a 0.9\,$\mu$m long active region measured at $H$ near 1.5\,kOe, $\phi=70^\circ$ and $I_{\mathrm{dc}}=1$\,mA by different techniques. \textbf{ b}~Conventional technique without field modulation. \textbf{ c}~Field modulation technique at fixed magnetic field and swept frequency. \textbf{ d}~Field modulation technique at fixed detection frequency and swept magnetic field. \textbf{ e}~ Spectrum from \textbf{ d} integrated with respect to the magnetic field.}
\label{DirectComp}
\end{figure*}

\begin{figure} 
\includegraphics[width=0.45\textwidth]{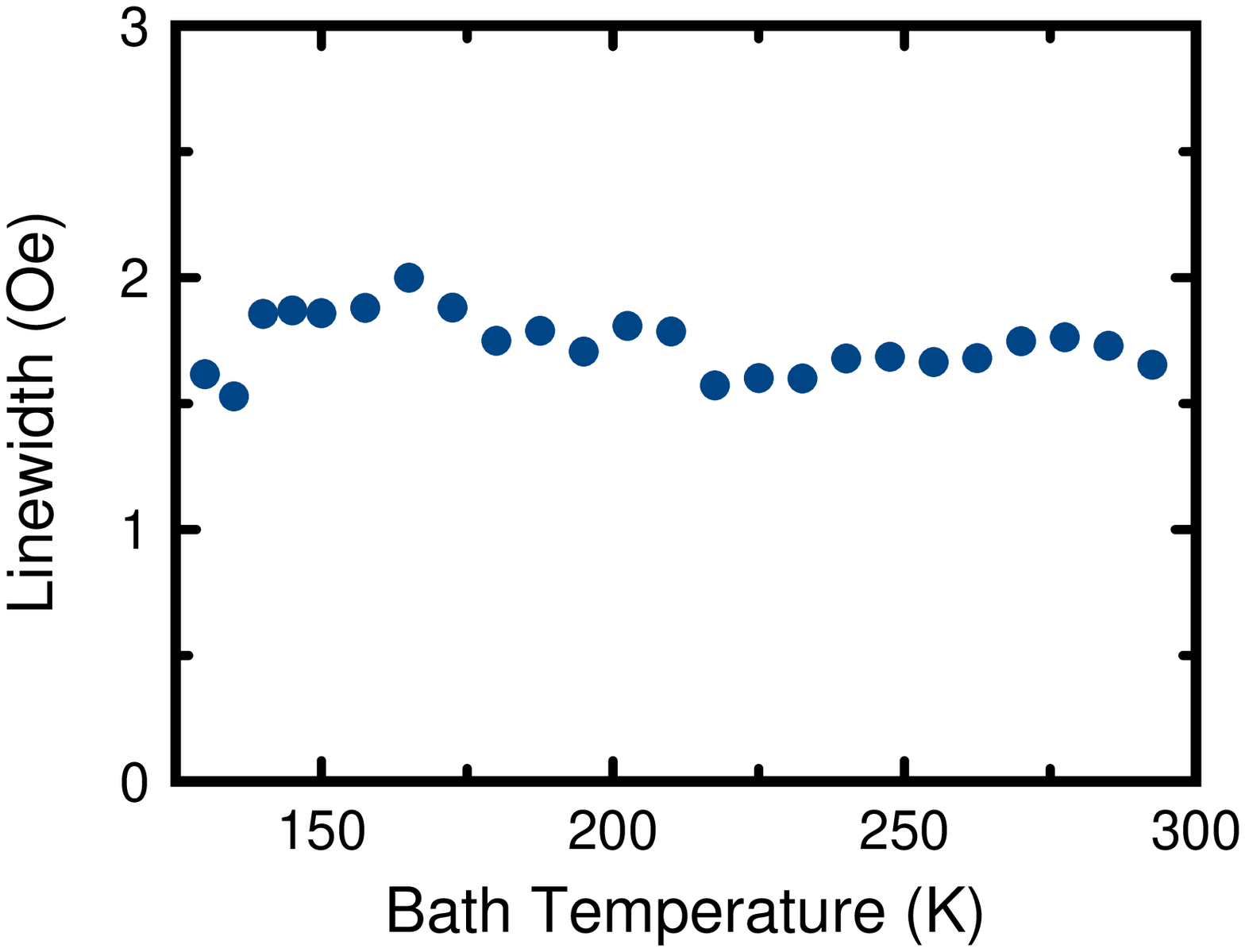}
\caption{\textbf{Temperature dependence of linewidth.} ST-FMR linewidth of the LF$_1$ mode as a function of bath temperature measured at the field angle $\phi=15^\circ$, drive frequency of 3.2\,GHz and drive power $P_{\mathrm{rf}}=1$\,dBm.}
\label{FMR_T}
\end{figure}

\section*{Supplementary Note 1. Ohmic Heating.} %Ohmic heating

In order to estimate the magnitude of thermal gradients arising from ohmic heating, we carried out finite element simulations of coupled electrical and thermal transport in the YIG/Pt nanowire devices using COMSOL Multiphysics package \cite{comsol}. We employed a fully realistic three-dimensional nanowire device geometry illustrated in Supplementary Figures\,\ref{DevPat} and \,\ref{1}a. A detailed sketch of the YIG/Pt nanowire device with two electrical leads used in the experiment and simulations is shown in Supplementary Figure\,\ref{DevPat}. The device consists of a 350\,nm wide 15\,$\mu$m long YIG(23\,nm)/Pt(8\,nm) nanowire with two Al(4\,nm)/Pt(2\,nm)/Cu(15\,nm)/Pt(2\,nm) electrical leads attached to the wire ends as shown in Supplementary Figure\,\ref{DevPat}. The leads are separated by a 2.5\,$\mu$m gap that defines the active region of the nanowire STO device.

In these simulations, we used the temperature-dependent heat conductivities and heat capacities of YIG and GGG as reported in Ref.\,\onlinecite{YIG_GGG_data}. The Pt layer resistivity in our YIG(23\,nm)/Pt(8\,nm) nanowire devices was measured in the temperature range from 140 K to 300 K and found to be linear as expected: $\rho(T) = \rho_0 (1+\alpha T)$ with $\rho_0 = 3.35\times 10^{-7} \Omega\cdot$m and $\alpha=1.59\times 10^{-3} $K$^{-1}$, which is  similar to previously reported values in thin Pt films \cite{Pt_film_res,Pt_film_alpha}. The temperature-dependent heat capacity of Pt reported in Ref.\,\onlinecite{Pt_heat_capacity} was employed in the simulations, and the thermal conductivity of the Pt layer was calculated from its electrical conductivity via the Wiedemann-Franz law. Literature values of the thermal and electrical conductivity and heat capacity of the lead materials were employed \cite{comsol}.

Supplementary Figure\,\ref{1}a shows the calculated spatial distribution of temperature in the YIG/Pt nanowire device studied in this work at the bath temperature of 140\,K and direct current bias $I_{\mathrm{dc}}=$ 2.5\,mA that is similar to the critical current. Supplementary Figures\,\ref{1}b and \ref{1}c show the depth profiles of the temperature in the center of the  YIG/Pt wire. These figures reveal that the Pt layer temperature rises to 220\,K.  The temperature in the YIG layer rapidly decreases with depth resulting in a large temperature gradient $\nabla T=0.26$\,K\,nm$^{-1}$ across the YIG layer thickness. This high degree of ohmic heating and the large value of $\nabla T$ result from the high resistivity of the Pt layer and efficient heat channeling into the YIG underlayer in the nanowire geometry employed in our experiment.  In this geometry, the metallic leads do not function as efficient heat sinks because their overlap area with the nanowire is relatively small, which results in a high degree of ohmic heating of the Pt nanowire and dissipation of this heat is mainly through the GGG/YIG underlayers. The quasi-one-dimensional nature of the Pt nanowire heat source and the three-dimensional character of the heat flow in the GGG substrate further enhance $\nabla T$ across the thickness of the YIG layer.

The validity of these COMSOL simulations can be directly checked against the experiment because the temperature of the Pt wire can be determined by measuring its resistance. Supplementary Figure \ref{2} shows the resistance of the Pt nanowire measured as a function of direct current bias. These data and the linear relation between the Pt nanowire resistance and temperature reveal that the Pt nanowire temperature at the bath temperature of 140\,K and $I_{\mathrm{dc}}=$ 2.5\,mA is 260\,K. We therefore conclude that the COMSOL simulations underestimate the degree of ohmic heating of the Pt wire and that the actual temperature gradient across the YIG film thickness is likely to exceed that predicted by the simulations.

We also employed COMSOL simulations to evaluate the ohmic heating in the YIG/Pt microdisk spin torque oscillators investigated in Ref.\,\onlinecite{Collet2016}. The microdisk device geometry used in our simulations and experimentally studied in Ref.\,\onlinecite{Collet2016} is shown in Supplementary Figure\,\ref{1}d. The system consists of a 2\,$\mu$m-diameter YIG(20\,nm)/Pt(8\,nm) disk defined on top of a GGG substrate. Two Ti(20\,nm)/Au(80\,nm) leads are attached to the disk with the inter-lead gap of 1\,$\mu$m. The system is covered with an SiO$_2$(300\,nm)/Au(250\,nm)  bilayer (not shown in Supplementary Figure\,\ref{1}d).  The Pt layer resistivity $\rho = 1.7\times 10^{-7} \Omega\cdot$m was directly measured for this system \cite{Collet2016}, and thermal conductivity of Pt was calculated via the Wiedemann-Franz law. In these simulations we also use the experimental parameters of Ref.\,\onlinecite{Collet2016}: a bath temperature of 293\,K and a critical current of 7.4\,mA. Literature values of the temperature-dependent electrical conductivity, thermal conductivity and heat capacity of Au, Ti, SiO$_2$, YIG and GGG \cite{Pt_heat_capacity,YIG_GGG_data,comsol} were employed in the simulations.

Supplementary Figures\,\ref{1}e,f show the depth dependence of the temperature in the center of the disk at $I_{\mathrm{dc}}=$ 7.4\,mA. It is clear from these figures that ohmic heating of the Pt layer is substantially smaller than  in our nanowire devices due to the lower electrical resistivity of the Pt layer employed in  Ref.\,\onlinecite{Collet2016} and better heat sinking by the Ti/Au leads having significant contact area with the microdisk. Combined with this, these devices are encased by 300\,nm of silicon oxide to allow electrical isolation from a 250 \,nm thick Au antenna, providing further heat sinking. The resulting temperature gradient in the YIG film across its thickness at the critical current is only 0.033\,K\,nm$^{-1}$ -- an order of magnitude smaller than that in our nanowire devices. Therefore, it is not surprising that the antidamping torque in these devices predominantly arises from spin Hall current with a negligible contribution from spin Seebeck current as evidenced by the $1/\sin\phi$ angular dependence of the critical current observed for these samples \cite{Collet2016}.

\section*{Supplementary note 2. Field modulated detection of microwave emission.}

Supplementary Figure\,\ref{DirectComp}a schematically illustrates the experimental setup employed in our field-modulated microwave emission measurements. A low-frequency ($\sim$1\,kHz), small-amplitude harmonic magnetic field $H_{\mathrm{ac}}$ is applied to an STO sample parallel to a constant external magnetic field $H_{\mathrm{dc}}$. A direct current bias $I_{\mathrm{dc}}$ is supplied from a custom built low noise current source and applied to the STO via a Picosecond 5541A-104 bias tee. The direct bias current excites the self-oscillations of magnetization. The microwave signal emitted by the sample is then amplified through a Miteq AMF-6F-00100400-10-10P low noise microwave amplifier with 62 dB gain, noise figure of 1\,dB, and frequency band of 0.1--4\,GHz. The amplified signal is then sent to an Agilent E4408B spectrum analyzer, configured in a single-frequency continuous detection mode. This configuration measures the integrated microwave power  in a 5\,MHz bandwidth around a fixed measurement frequency. The STO generation frequency is modulated by $H_{\mathrm{ac}}$, which results in a modulation of the STO power at the measurement frequency. The modulated STO emission power $\frac{\mathrm{d}P}{\mathrm{d}H}$ is measured by a Signal Recovery 7225 lock-in amplifier via a video  output port of the spectrum analyzer. In order to obtain a field-modulated STO emission spectrum, the data is collected point-by-point by stepping the measurement frequency of the spectrum analyzer over a desired frequency range.

The conventional method of measuring STO microwave emission spectrum, in which the emission power is simply recorded as a function of frequency, did not yield a signal exceeding the noise floor for the 350\,nm wide YIG/Pt nanowire samples discussed, and our field modulation technique was required to observe the signal.  In order to quantitatively compare the field-modulated emission method to the conventional method, we employ an STO sample based on a 90\,nm wide YIG(23\,nm)/Pt(8\,nm) nanowire with a 0.9\,$\mu$m long active region. This STO generates higher microwave signals, which can be measured by the conventional technique as illustrated in Supplementary Figure\,\ref{DirectComp}b. 

Supplementary Figures\,\ref{DirectComp}b,c directly compare the microwave emission spectra for conventional and field-modulated detection measured under identical conditions ($H=1.5$\,kOe, $\phi=70^\circ$, $I_{\mathrm{dc}}=1$\,mA, measurement time 17 minutes). The conventional method gives a spectral peak with integrated power of 21\,fW.  In contrast, the field modulation method yields a prominent $\frac{\mathrm{d}P}{\mathrm{d}H}$ signal with high signal-to-noise ratio and the line shape similar to a Lorentzian curve derivative as illustrated in Supplementary Figure\,\ref{DirectComp}c. 

Supplementary Figures\,\ref{DirectComp}d,e illustrate that the field modulation method can be further improved by sweeping external magnetic field instead of stepping the center-frequency as done in Supplementary Figure\,\ref{DirectComp}c. In Supplementary Figure\,\ref{DirectComp}d, the field-modulated emission signal is measured as a function of applied field giving the expected antisymmetric line shape. This signal can be directly integrated in magnetic field yielding a symmetric emission curve as a function of magnetic field. The development of this microwave detection technique increases the signal-to-noise ratio by two orders of magnitude, allowing detection of ultra low-level microwave signals emitted by magnetic devices.

Approximate calibration of the power scale for the field-integrated spectra such as that shown in Supplementary Figure\,\ref{DirectComp}e can be performed via comparison of the spectral peak amplitudes in Supplementary Figures\,\ref{DirectComp}b,e. We estimate the maximum power spectral density $P_{\mathrm{max}}$ generated by the 350 nm wide YIG/Pt nanowire shown in Fig.\,2b to be approximately 0.1\,fW\,MHz$^{-1}$ and the corresponding integrated power to be approximately 6\,fW.

We also note that the nanowire geometry can be used to tune both the frequency and the amplitude of the microwave signal generated by the YIG/Pt nanowire STO. We find that decreasing the width of the nanowire from 350\,nm to 90\,nm results in a decrease of the generated signal frequency. At the same time, the output power of the STO increases by over a factor of three. The decrease of the resonance frequency results from a higher demagnetization field in the narrower nanowire \cite{Duan2015} while the increase of the output power can be attributed to a larger volume fraction of the wire occupied by the spin wave mode (see Fig.\,4c).

\section*{Supplementary Note 3. Auto-oscillation amplitude.}

The precession cone angle of the auto-oscillatory YIG magnetization  can be estimated from the output microwave power of the YIG/Pt nanowire STO. The integrated microwave power $P_\mathrm{{int}}$ generated by an STO is proportional to the square of the direct current bias $I_\mathrm{dc}$ and the amplitude of resistance auto-oscillations $\delta R_\mathrm{ac}$  \cite{Kiselev2003,Duan2014}:
\begin{eqnarray}
{ P_{\mathrm{int}} =  \dfrac{1}{2 R_{50}} \left ( I_{dc} \delta R_{ac} \dfrac{R_{50}}{ R+R_{50}}    \right )^2  } 
\label{Ep}\end{eqnarray}
where $R$ is the sample resistance and $R_\mathrm{50}$ is the 50\,$\Omega$ microwave transmission line impedance. Assuming the angular dependence of the YIG/Pt nanowire resistance is $R = R_0+\frac{{\Delta R}}{2} \cos 2\phi$ as expected for SMR, the small-amplitude dynamic resistance oscillations $\delta R_\mathrm{ac}$ are related to the in-plane precession cone angle $\phi_{\mathrm{c}}$ in the macrospin approximation as: 
\begin{eqnarray}
{ \delta R_{ac} =  \phi_{c}  \Delta \noindent R \sin2 \phi_0} 
\label{Ep2}\end{eqnarray}
where $\phi_0$ is the equilibrium direction of the YIG magnetization. The maximum value of $\phi_{\mathrm{c}}$ achieved by the YIG magnetization in the 350\,nm wide nanowire device can be calculated from Equations 1 and 2 by using the  generated integrated power $P_\mathrm{{int}}=$ 6\,fW and $\Delta R=$ 0.05\,$\Omega$ extracted from Fig.\,1b. This calculation gives the precession cone angle in the macrospin approximation $\phi_{\mathrm{c}}\approx$ 6$^\circ$. Taking into account that the excited LF mode has the edge character as shown in Fig.\,4c, and that the edge mode occupies approximately one third of the nanowire volume as predicted by our micromagnetic simulations, the amplitude of the YIG magnetization oscillations at the nanowire edge is estimated to be approximately $20^\circ$. We stress that this is merely an estimate because contributions to the generated microwave signal beyond SMR such as inductive signal generated by precessing magnetization \cite{Collet2016} can be non-negligible in our nanowire devices.

\section*{Supplementary Note 4. Micromagnetic simulations}
Micromagnetic simulations of the spin wave eigenmode frequencies of the YIG/Pt nanowires were performed using a modified version of the finite-differences simulation code MuMax\textsuperscript{3} \cite{MuMax3_2014}. The nanowire was discretized
into $2048\times32\times4$ cells, resulting in a cell size of $6.30\times8.75\times7.50$\,nm\textsuperscript{3}. The saturation magnetization $M_{\mathrm{s}}= 130$\,kA\,m$^{-1}$ \cite{Solt1962} and the exchange constant
$A_{\mathrm{ex}}=3.5$\,pJ\,m$^{-1}$ were used \cite{klingler2014}. The spin wave eigenfrequencies were determined as the peak position of the  Fourier-transform of the dynamic magnetization excited by a sinc-shaped magnetic field pulse \cite{Venkat2013}. The simulation time was chosen to be 25\,ns, which results in an FFT frequency resolution of 40\,MHz.  The spin wave profiles shown in Fig.\,4c are represented by the cell-specific Fourier amplitude.

\section*{Supplementary Note 5. Magnetic damping.}

We use the conventional broadband FMR technique \cite{Sangita} to measure the FMR linewidth $\Delta H$ (defined as half width at half maximum of the Lorentzian absorption curve) as a function of frequency $f$ for YIG(23\,nm) and YIG(23\,nm)/Pt(8\,nm) films. The damping constant $\alpha$ and the inhomogeneous broadening parameter $\Delta H_0$ are determined from the slope and zero-frequency intercept of the $\Delta H(f)$ data \cite{FarleFMR}. These measurements give $\alpha=0.0014$ and $\Delta H_0= 1.1$\,Oe for the YIG film and $\alpha = 0.0035$ and $\Delta H_0 = 2.6$\,Oe for the YIG/Pt bilayer. The linewidth $\Delta H$ measured at 3.2\,GHz for the YIG/Pt bilayer film is found to be 6.7\,Oe, which is similar to $\Delta H$ = 6\,Oe at 3.2\,GHz measured in the YIG/Pt nanowire device by ST-FMR at the lowest microwave power value ($-5$\, dBm) as shown in Fig.\,4e. This demonstrates that patterning of the YIG/Pt film into the nanowire device does not significantly alter the YIG layer damping.

\section*{Supplementary Note 6. Temperature dependence of linewidth.}

Ohmic heating of Pt can affect magnetization dynamics in the YIG/Pt nanowire via both the thermal gradient across the YIG/Pt interface and the increase of the average temperature of YIG. To separate these effects, we carried out ST-FMR measurements as a function of bath temperature at a constant low direct current bias (0.15\,mA). Supplementary Figure\,\ref{FMR_T} shows temperature dependence of ST-FMR linewidth of the LF$_1$ mode measured at $\phi=15^\circ$ in a YIG(23\,nm)/Pt(8\,nm) nanowire device nominally identical to that shown in Fig.\,4. The measurements reveal that the linewidth is nearly constant in the 140\,K -- 300\,K temperature range. The data in Supplementary Figure\,\ref{FMR_T} demonstrate that the current-induced linewidth variation in Fig.\,4g arises from spin Seebeck current driven by $\nabla T$ and not from temperature dependence of the YIG nanowire damping.

\end{document}